\documentclass[aps,prl, preprint]{revtex4-1}
\usepackage{amsfonts}
\usepackage{amsmath}
\usepackage{amssymb}
\usepackage{graphicx}
\usepackage{times}
\usepackage{physics}
\usepackage[colorlinks=true,linkcolor=blue,citecolor=red,plainpages=false,pdfpagelabels]
{hyperref}
\usepackage[encapsulated]{CJK}

\usepackage{color}

\providecommand{\U}[1]{\protect\rule{.1in}{.1in}}

\begin{document}

\title{Extract the Degradation Information in Squeezed States  with Machine Learning}
\author{Hsien-Yi Hsieh,$^{1}$ Yi-Ru Chen,$^{1}$ Hsun-Chung Wu,$^{1}$ Hua Li Chen,$^{2}$ Jingyu Ning,$^{1}$  Yao-Chin Huang,$^{1}$ Chien-Ming Wu,$^{1}$ and Ray-Kuang Lee$^{1,2,3,4}$}
\affiliation{$^{1}$Institute of Photonics Technologies, National Tsing Hua University, Hsinchu 30013, Taiwan\\
$^{2}$Department of Physics, National Tsing Hua University, Hsinchu 30013, Taiwan\\
$^{3}$Physics Division, National Center for Theoretical Sciences, Taipei 10617, Taiwan\\
$^{4}$Center for Quantum Technology, Hsinchu 30013, Taiwan}
 \email{rklee@ee.nthu.edu.tw}

\date{\today}
\begin{abstract}
In order to leverage the full power of quantum noise squeezing with unavoidable decoherence, a complete understanding of the degradation in the purity of  squeezed light is demanded.
By implementing machine learning architecture with a convolutional neural network, we illustrate a fast, robust, and precise quantum state tomography for continuous variables, through the experimentally measured data generated from the balanced homodyne detectors. 
Compared with the maximum likelihood estimation method,  which suffers from time-consuming and over-fitting problems, a well-trained machine fed with  squeezed vacuum  and squeezed thermal states can complete the task of reconstruction of the density matrix in less than one second.
Moreover, the resulting fidelity remains as high as $0.99$ even when the anti-squeezing level is higher than $20$~dB.
Compared with the phase noise and loss mechanisms coupled from the environment and surrounding vacuum, experimentally, the degradation information is unveiled with machine learning  for  low and high noisy scenarios, i.e., with the anti-squeezing levels at $12$~dB and $18$~dB, respectively.
Our neural network enhanced quantum state tomography provides the metrics to give physical descriptions of every feature observed in the quantum state with a single-shot measurement and paves a way of exploring large-scale quantum systems in real-time. 
\end{abstract}

\maketitle

With the intrinsic nature of  multimode, continuous variable states have provided a powerful platform for generating large entangled networks~\cite{CV, CV-cluster, 60, 1M, 2Dcluster-1, 2Dcluster-2}.
In the family of continuous variables, {\it squeezed states},  even with the fundamental limit on the quantum fluctuations set by  Heisenberg’s uncertainty relation, remarkably  exhibit completely different characteristics in the quantum world~\cite{Yuen, SQ-book,  SQ-30}.
Now as true applications, squeezed states have been used in quantum metrology~\cite{metro-1, metro-Science, metro-NP, metro-NP2}, advanced gravitational wave detectors ~\cite{Cave, GW-13, GW-18, GW-19, GW-Virgo, GW-FDSQ-1, GW-FDSQ-2}, generation of macroscopical states~\cite{cat1, cat2, cat3}, and quantum information manipulation utilizing continuous variables~\cite{CV-teleportation, Gaussian}.

Even though up to $15$~dB squeezing has been demonstrated as the state-of-the-art technology, any quantum system is unavoidably subject to a number of dissipative processes, causing $18$-$24$~dB anti-squeezing accompanied~\cite{15dB}.
Instead of dealing with pure states, the degradations in squeezing from  loss and phase noise fluctuations limit the practical applications, resulting in tackling with mixed states.
The imperfection in purity is not only the obstacle for any quantum metrology with squeezed states, but also the restriction in generating larger-size  Schr{\" o}dinger's cat states.
To access the non-classical power for quantum technologies,  we need to have the ability to fully and precisely characterize the quantum features in a large Hilbert space. 
Utilizing of multiple phase-sensitive measurements through homodyne detectors, quantum state tomography (QST) enables us to extract the complete information about the state of the system statistically~\cite{Hradil, QST}.
Nowadays, QST has been implemented in a variety of  quantum systems, including quantum optics~\cite{QST-book,QST-Furusawa}, ultracold atoms~\cite{QST-atom1, QST-atom2}, ions~\cite{QST-ion-1, QST-ion-2}, and  superconducting circuit-QED devices~\cite{QST-SC}. 

\begin{figure}[b]
\includegraphics[width=8.6cm]{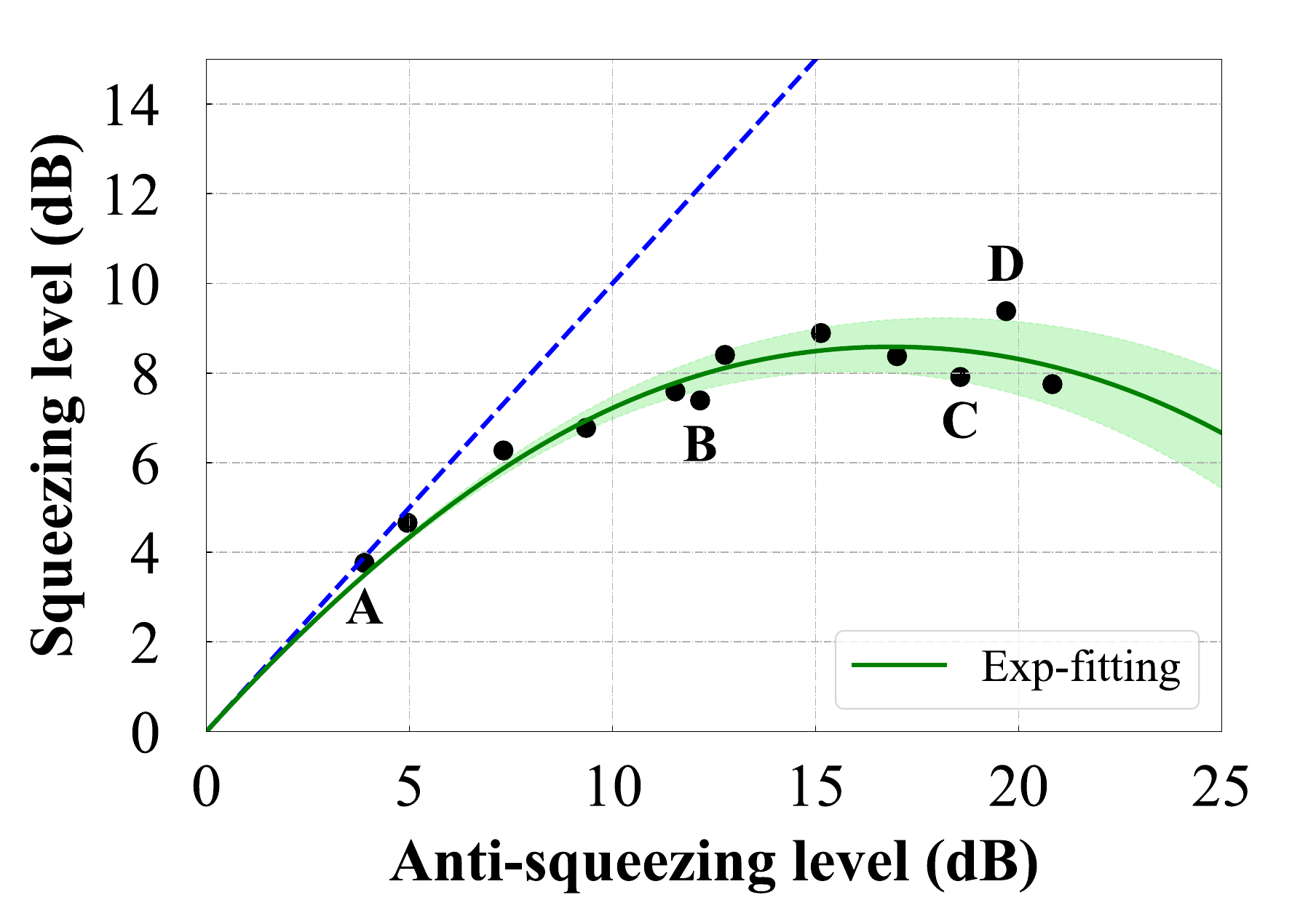}
\caption{Degradation in squeezed states. Ideally, the squeezing and anti-squeezing levels should locate along the Blue-dashed line. However, as shown with the typical experimental data, marked in Black dots, there exits a discrepancy between the measured squeezing and anti-squeezing levels. By taking the loss and phase noise into account, the optimal fitting curve is depicted in Green-color, with the corresponding standard deviation shown by the shadowed region. }
\end{figure}

\begin{figure*}[ht]
\begin{flushleft}
\includegraphics[width=18cm]{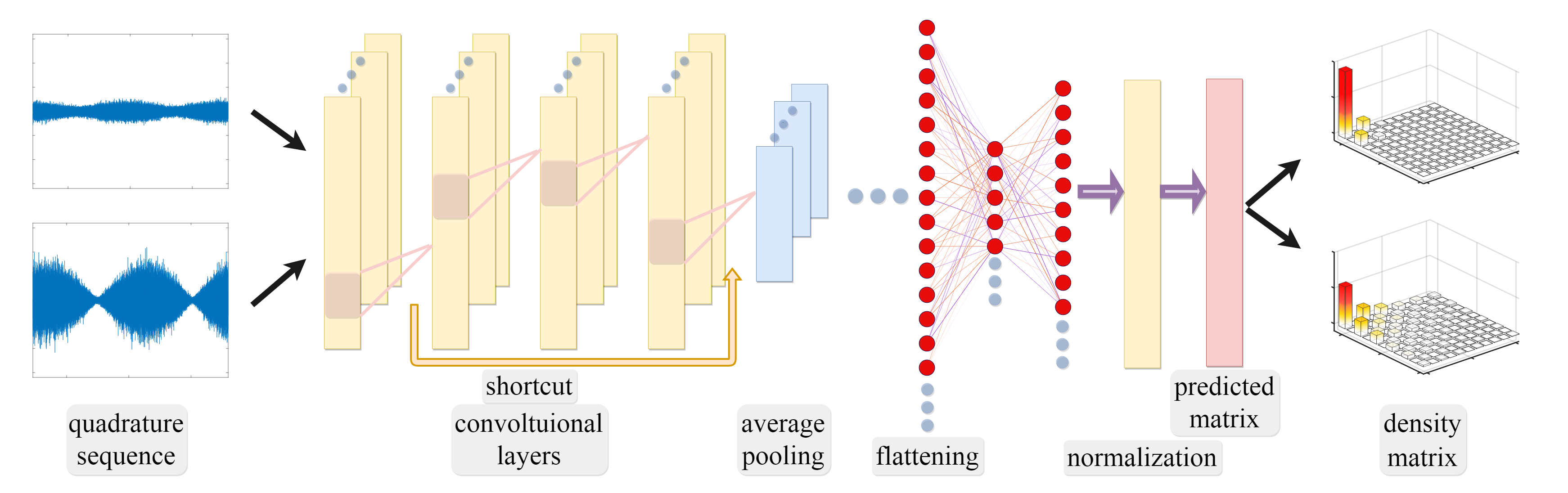}
\caption{Schematic of our precise and robust neural network enhanced quantum state tomography. The noisy data of quadrature sequence obtained by quantum homodyne tomography are fed to the convolutional layers, with the shortcut and average pooling in the architecture. Then, after flattening and normalization, the predicted matrices are inverted to reconstruct the density matrices in truncation.}
\label{Fig1}
\end{flushleft}
\end{figure*}

One of the most popular methods to implement QST is the maximum likelihood estimation (MLE) method, by estimating the closest probability distribution to the data for any arbitrary quantum states ~\cite{MLE}. 
However, the required amount of measurements to reconstruct the quantum state in multiple bases  increases exponentially with the number of involved modes.
Albeit  dealing with Gaussian quantum states,   the MLE algorithm becomes computationally too heavy and intractable when the squeezing level increases. 
Moreover, MLE also suffers from the over-fitting problem when the number of bases grows. 
To make QST more accessible, several alternative algorithms are proposed by assuming some physical restrictions imposed upon the state in question, such as the permutationally invariant tomography~\cite{permu}, quantum compressed sensing~\cite{compress}, tensor networks~\cite{tensor-1, tensor-2}, and generative models~\cite{generative}.
Instead, with the capability  to find the best fit to arbitrarily complicated data patterns with a limited number of parameters available,   machine learning approaches are widely applied in many sub-fields in physics, from black hole detection, topological codes,  phase transition, to quantum physics~\cite{QML-1, QML-2}. 
For QST,  the restricted Boltzmann machine has been applied to reduce the over-fitting problem in MLE~\cite{RBM}. 

 In dealing with  Gaussian states, methodologies based on  covariance matrix or nullifiers are well-developed~\cite{witness, comb, nullifier}.  
However, more than one single measurement is needed for optical homodyne tomography.
Moreover, the assumption of Gaussian properties in the generated squeezed state is only valid for low squeezing levels. 
When the squeezing level is higher than $5$ dB, more and more non-squeezing parts become dominant, making these known methodologies inaccurate.
Nowadays, higher than $10$ dB squeezing levels are  in the schedule for the advanced gravitational wave detectors~\cite{LIGO-quantum, Virgo}. 
Even though for the non-ideal case, one can also apply the nullifiers to represent the actual noises in the operations  by additional feedforward operations.
A single-shot measurement to extract the degradation in quantum states  is still missing.

Along this direction,  based on the machine learning protocol,  in particular with the convolutional neural network (CNN), we experimentally implement the quantum homodyne tomography for continuous variables and illustrate a fast, robust, and precise QST for squeezed states. 
As the time sequence data obtained in the optical homodyne measurements share the similarity to the voice (sound) pattern recognition, it motivates us to apply the CNN architecture.
With the aim of realizing a fast QST,  such a supervised CNN trained by the prior knowledge in squeezed states enables us to build a  specific machine-learning for certain kinds of problems.
More than two million data sets are fed into our machine with a variety of squeezed  and thermal states in different squeezing levels,  quadrature angles, and reservoir temperatures.
Compared with the time-consuming MLE method, demonstrations on the reconstruction of the Wigner function and the corresponding density matrix are illustrated for squeezed vacuum states in less than one second, keeping the fidelity up to $0.99$ even  taking $20$~dB anti-squeezing level into  consideration.
Experimentally, the purity in squeezed vacuum states is evaluated directly for high squeezing levels (close to $10$ dB squeezing level), but with  low noisy and high noisy conditions, i.e., with the anti-squeezing levels at $12$~dB and $18$~dB, respectively.
By extracting the purity of quantum states with the help of machine learning, a full understanding of the degradation in the state decoherence  can also be  unveiled in a single shot measurement, paving the road toward a real-time QST to give physical descriptions of every feature observed in the quantum noise.

First of all,  in Fig. 1, we show the  degradation curve in  typical squeezed state experiments, illustrated with the measured squeezing and anti-squeezing levels in the unit of decibel (dB).
Here, our squeezed vacuum states are generated through a bow-tie optical parametric oscillator cavity with a periodically poled KTiOPO$_4$ (PPKTP) inside, operated below the threshold at the wavelength $1064$ nm~\cite{CLEO}. 
By injecting the AC signal of our homemade balanced homodyne detection, with the common-mode rejection ratio (CMRR) more than $80$~dB, the spectrum analyzer records the squeezing and anti-squeezing levels by scanning the  phase of the local oscillator.
Specifically, four experimental data are marked with the measured (squeezing:anti-squeezing) levels in dB: $A$ ($3.76$: $3.89$) at the pump power $5$ mW, $B$ ($7.39$:$12.16$) at $55$ mW, $C$ ($7.91$: $18.56$) at $77$ mW, and $D$ ($9.38$:$19.69$) at $80$ mW, respectively.

Ideally, without any degradation, the squeezing and anti-squeezing levels should be the same, located along the Blue-dashed line in Fig. 1.
However, the phase noise and loss mechanisms coupled from the environment and surrounding vacuum set the limit on the measured squeezing level.
These selected data represent nearly ideal squeezing (marker $A$), high squeezing level but with low degradation (marker $B$) and with high degradation (marker $C$); along with the highest squeezing level achieved (marker $D$ with $9.38$ dB in squeezing level).
Empirically, quantum radiation pressure noise (QRPN) is known as the most prominent in the observation from squeezing~\cite{QRPN-1, QRPN-2}. 
 By taking the optical loss (denoted as $L$) and phase noise (denoted as $\theta$) into account, the measured squeezing $V^{sq}$ and anti-squeezing $V^{as}$ levels can be modeled as
\begin{eqnarray}
&&V^{sq} = (1-L) [V^{sq}_{id}\times \cos^2 \theta + V^{as}_{id}\times \sin^2\theta]+L,\\
&&V^{as} = (1-L)[V^{as}_{id}\times \cos^2\theta + V^{sq}_{id}\times \sin^2\theta]+L,
\end{eqnarray}
where $V^{sq}_{id}$ and $V^{as}_{id}$ are the squeezing and anti-squeezing levels in the ideal case.
In Fig. 1, we also show the optimal fitting curve obtained by the orthogonal distance regression  in Green-color, with the corresponding  standard deviation (one-sigma variance) shown by the shadowed region.

\begin{figure}[t]
\includegraphics[width=8.6cm]{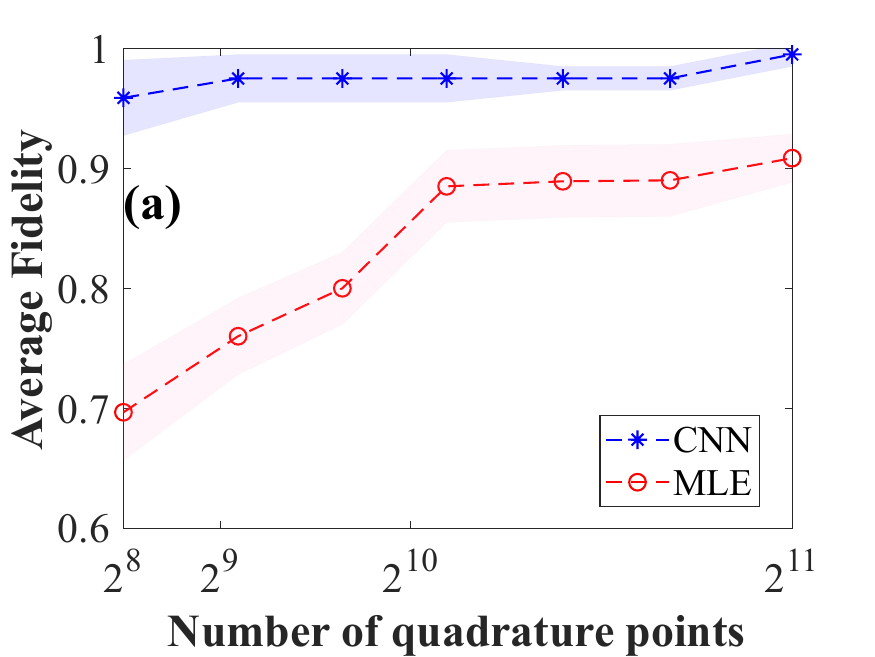}
\includegraphics[width=8.6cm]{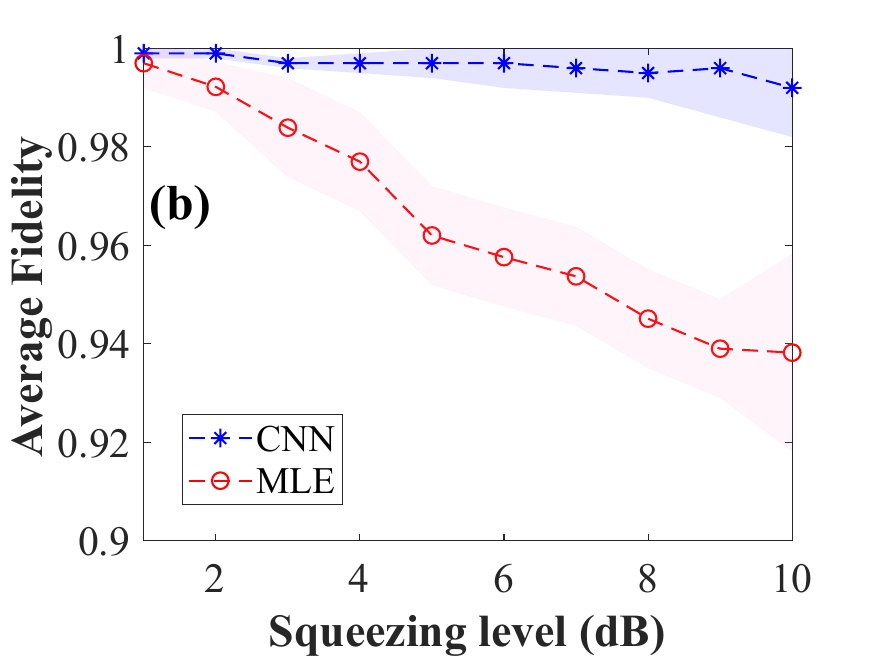}
\caption{Comparison in the average fidelity for the predicted density matrix obtained by maximum likelihood estimation (MLE) and convolution neural network (CNN) as a function of  (a) the number of quadrature data points, and (b) the squeezing level.
Here, $5,000$ simulated data are prepared for the comparison, but  in (a) with different squeezing levels from $8$ to $14$~dB; while in (b), the number of data points in the quadrature sequence is fixed to  $2,048$. The shadow regions represent the standard  deviation in the average fidelity.}
\end{figure}

Even though by fitting several measured squeezing and anti-squeezing data one can estimate the degradation in squeezing empirically, the full information about the density matrix and the purity of  quantum states needs to be reconstructed precisely and fast. 
To generate QST for continuous variables, keeping the  fidelity high and avoiding non-physical states are the critical issues in training  our neural network enhanced tomography scheme.
As illustrated in Fig. 2, the noisy data of quadrature sequence obtained by quantum homodyne tomography are fed into a convolutional neural network (CNN), composited with $30$ convolutional layers in total.
To extract the resulting density matrix from the time series data, i.e., the quadrature sequence, we take the advantage of  good generalizability in applying CNN~\cite{generalizability}.
There are four convolution blocks used in our deep CNN, each of which contains $1$ to $9$ convolution layers (filters) in different sizes.
 In order to tackle the gradient vanishing problem, which  commonly happens in the deep CNN when the number of convolution layers increases~\cite{GV}, five shortcuts are also introduced among the convolution blocks.
Instead of the max-pooling, the average pooling is applied to produce higher fidelity results, as all the tomography data should be equally weighted.
Finally, after flattening two fully connected layers  and normalization, the predicted matrices are inverted to reconstruct the density matrices in truncation.
Here, the loss function we want to minimize is the mean squared error (MSE); while the optimizer used for training is Adam. We take the batch size as $32$ in the training process. By this setting, the network is trained with $70$ epochs to decrease the loss (MSE) up to $5\times 10^{-6}$.
Practically, instead of an infinite sum on the photon number basis, we keep the sum in the probability up to $0.9999$ by truncating the photon number.
Here, the resulting density matrix is represented in photon number basis, which is  truncated to $35\times 35$   by considering the maximum anti-squeezing level up to $20$~dB.

As to avoid non-physical states, we impose the positive semi-definite constrain into the predicted density matrix.
Here, an auxiliary (lower triangular) matrix is introduced before generating the predicted factorized density matrix  through the Cholesky decomposition.
During  the training process, the normalization also ensures  the trace of the output density matrix kept as $1$. 
Moreover, more than two million (exactly, $2,019,600$) data sets  are prepared for training, including squeezed vacuum states with different squeezing levels and quadrature angles, i.e., $ \rho^{sq} =\hat{S}\rho_{0}\hat{S}^\dag$, alone with the  squeezed thermal states in different reservoir temperatures, i.e.,   $\rho^{sq}_{th} =\hat{S}\rho_{th}\hat{S}^\dag$~\cite{sq-thermal, sq-thermal-2, sq-thermal-3}. 
Here, $\hat{S} (\xi) = \text{exp}[\frac{1}{2} \xi^\ast \hat{a}^2 - \frac{1}{2} \xi \hat{a}^{\dag 2}]$ denotes the squeezing operator, with the squeezing parameter $\xi \equiv r\, \text{exp}(i \phi)$ characterized by the squeezing factor $r$ and the squeezing angle $\phi$.
Moreover, $\rho_0$ and $\rho_{th}$ correspond to the density matrix of vacuum state and thermal state at a given reservoir temperature, respectively.
All the training is carried out with the Python package {\it tensorflow.keras} performed in GPU  (Nvidia Titan RTX).

To illustrate that our neural network enhanced QST indeed keeps the fidelity in the predicted  density matrix, in Fig. 3, the average fidelity obtained by MLE and CNN are compared as a function of (a) the number of quadrature data points and (b) the squeezing levels (dB).
Here, the fidelity is defined as $| \text{tr}(\sqrt{\sqrt{\rho }\, \sigma\, \sqrt{\rho }})|^{2}$, with the given simulated input data $\rho$ and the predicted density matrix $\sigma$ obtained by MLE and CNN, respectively.
The average is done with $5,000$ simulated data set. 
As one can see in Fig. 3(a), when the number of quadrature points increases, from $256$ to $2,048$, the resulting average fidelity increases even when we consider higher squeezing levels, i.e., $8$ to $14$~dB. 
Nevertheless,  only with a small number of data points such as $256$, the output fidelity obtained by CNN can be much higher than $0.95$, compared to only $0.7$ by MLE.
Moreover, even the data point increases to $2,048$,   MLE only gives the average fidelity up to $0.91$, which is still much lower than $0.99$ obtained by CNN.
 
On the other hand,  when the data points are fixed to $2,048$,  the superiority in CNN  over MLE can also be clearly seen in Fig. 3(b), in particular at a higher squeezing level.
Now, as the squeezing level increases, the dimension in the reconstructed Hilbert space exponentially grows.
As a result, the average fidelity obtained by MLE decreases very quickly from $0.99$ at a low ($1$~dB) squeezing level to $0.94$ at a high ($10$~dB) squeezing level.
On the contrary, our well-trained machine learning can keep the output fidelity as high as $0.99$ even when $10$~dB squeezing is tested.

\begin{figure}[t]
\includegraphics[width=8.6cm]{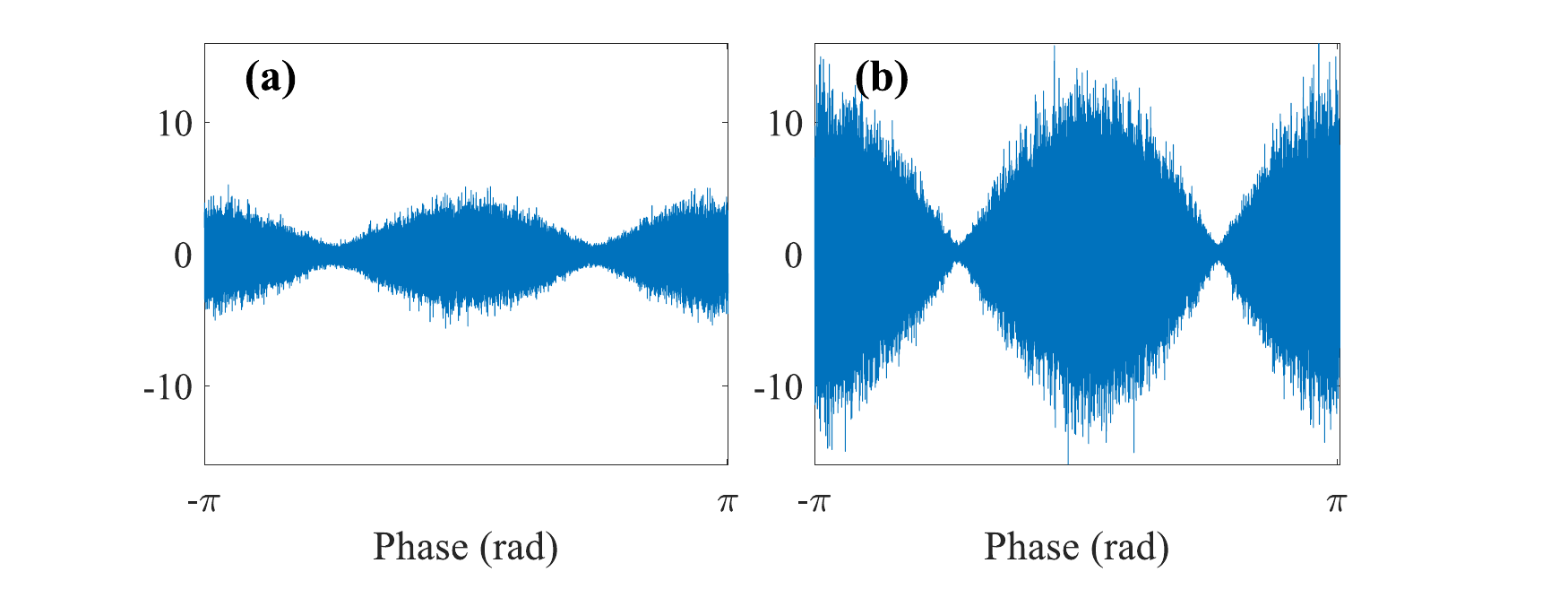}
\includegraphics[width=8.6cm]{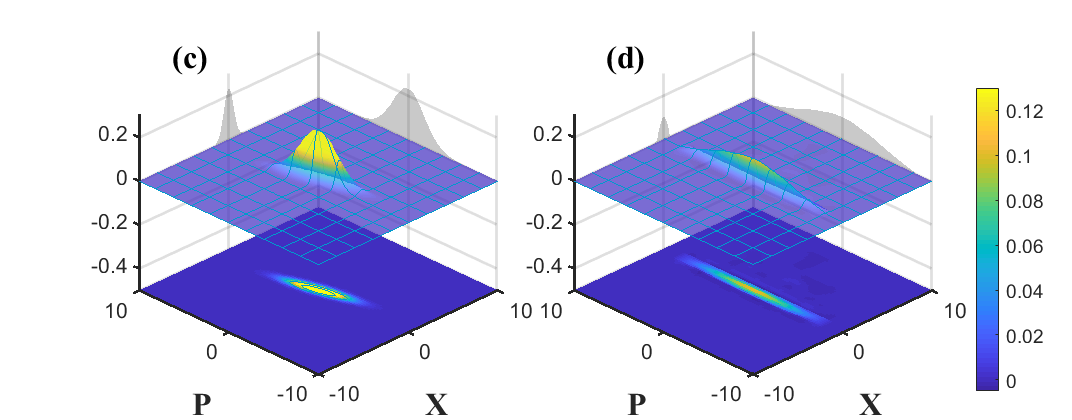}
\includegraphics[width=8.6cm]{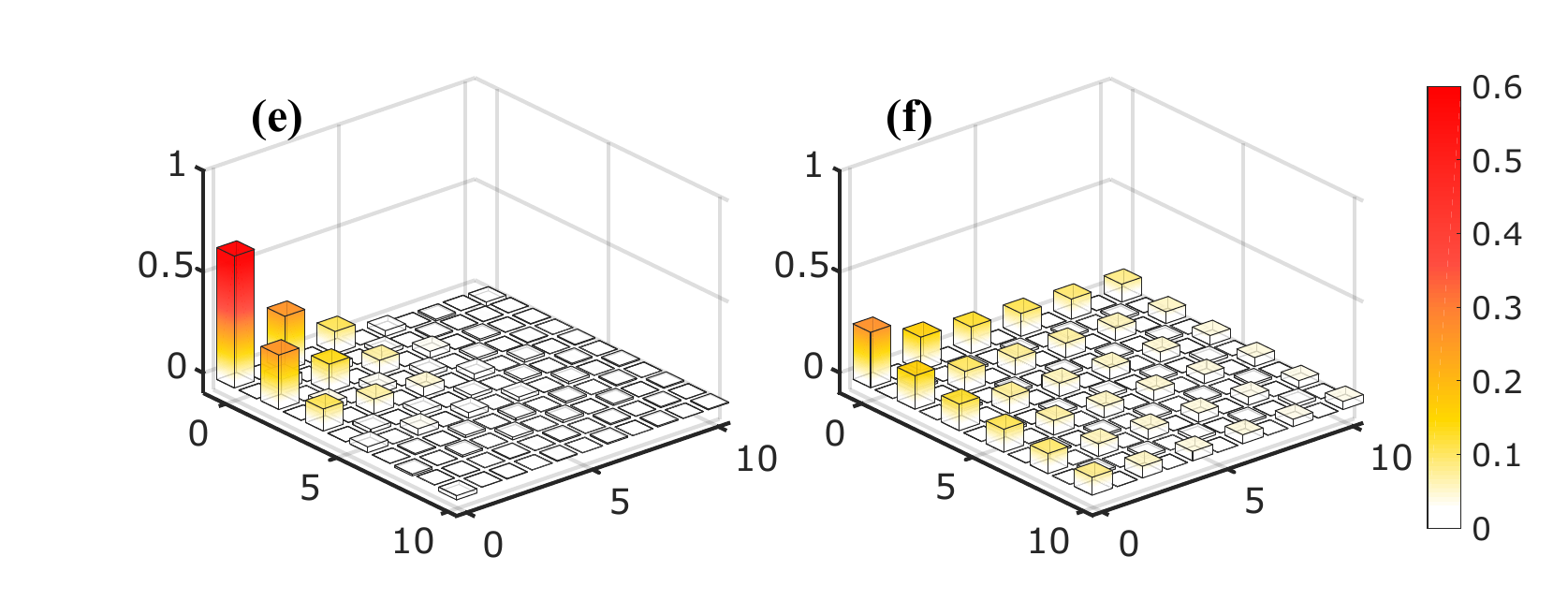}
\caption{The experimental raw data of squeezed vacuum states with high measured squeezing level ($>7$~dB), but (a) with low degradation ($12.16$~dB in the anti-squeezing), and (b) with high degradation ($18.56$~dB in the anti-squeezing), corresponding to the marker points $B$ and $C$ in Fig. 1.
 The predicted results by machine learning are shown for (c, d) the reconstructed Wigner functions $W(\mathbf{X}, \mathbf{P})$ and (e, f) the real part of the predict density matrix in the photon number basis, respectively. }
\end{figure}

Now, in Fig. 4, we apply this CNN-enhanced QST to our experimental raw data measured from the oscilloscope, with the quadrature discretized into  $2,048$ data points. 
In particular, as shown in Fig. 4(a) and (b), we focus on the high squeezing level cases, i.e., $>7$~dB, but compare with  low degradation ($12.16$~dB in the anti-squeezing) and high  degradation ($18.56$~dB in the anti-squeezing), corresponding to the marker points $B$ and $C$ shown in Fig. 1.
The corresponding Wigner functions $W(\mathbf{X},\mathbf{P})$ for quantum squeezed states in phase space are shown in Fig. 4(c) and (d), respectively.
As a comparison, the more degradation in squeezing, the reconstructed wave-package becomes broader.
The  predicted density matrixes in the photon number basis are depicted in Fig. 4 (e) and (f), accordingly.
Degradation information can be easily seen from the spread of density matrix elements.
Moreover, we want to remark that only with a single-shot measurement our well-trained machine  completes the task on the reconstruction of Wigner function and the corresponding density matrix in less than one second.

 \begin{figure}[t]
\includegraphics[width=8.6cm]{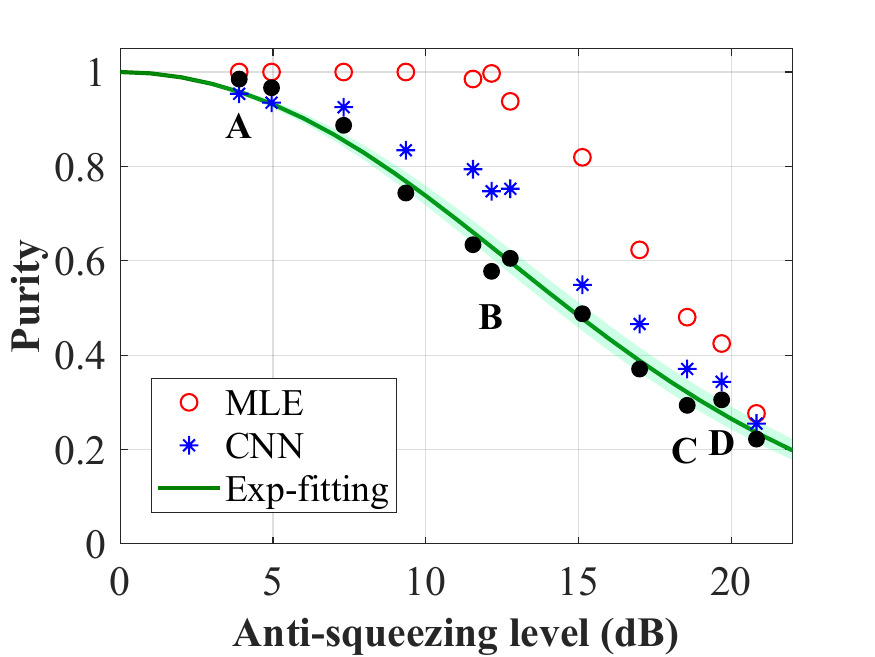}
\caption{The purity of squeezed states is plotted as a function of the measured anti-squeezing level. The experimental data marked in Fig. 1 are analyzed with MLE and CNN, plotted in  Red- and Blue-colors, respectively.
The fitting results based on Eqs. (1-2) are depicted in Green-curve, with the corresponding standard deviation shown in the shadow region. Here, at high squeezing levels, MLE over-estimates the purity of the quantum states in QST; while the empirical formula under-estimates the purity.}
\end{figure}

In addition to the reconstructed Wigner function and the corresponding density matrix, the degradation information in squeezing can be extracted directly from the predicted density matrix by calculating the purity of  quantum state, i.e., $p \equiv \text{tr}(\rho^2)$.
The performance of our machine-learning QST is compared with the one obtained by covariance matrix, denoted as the Exp-fitting curve in Fig. 5, as well as the one obtained by MLE, on the purity of squeezed states through the predicted density matrix.
In addition to exhibiting the same trend in the degradation of purity, as one can see, at high squeezing levels, MLE over-estimates the purity of quantum states due to the over-fitting problem.
On the contrary,  the empirical formula under-estimates the purity due to the lack of thermal reservoir information.

\begin{figure}[t]
\includegraphics[width=8.6cm]{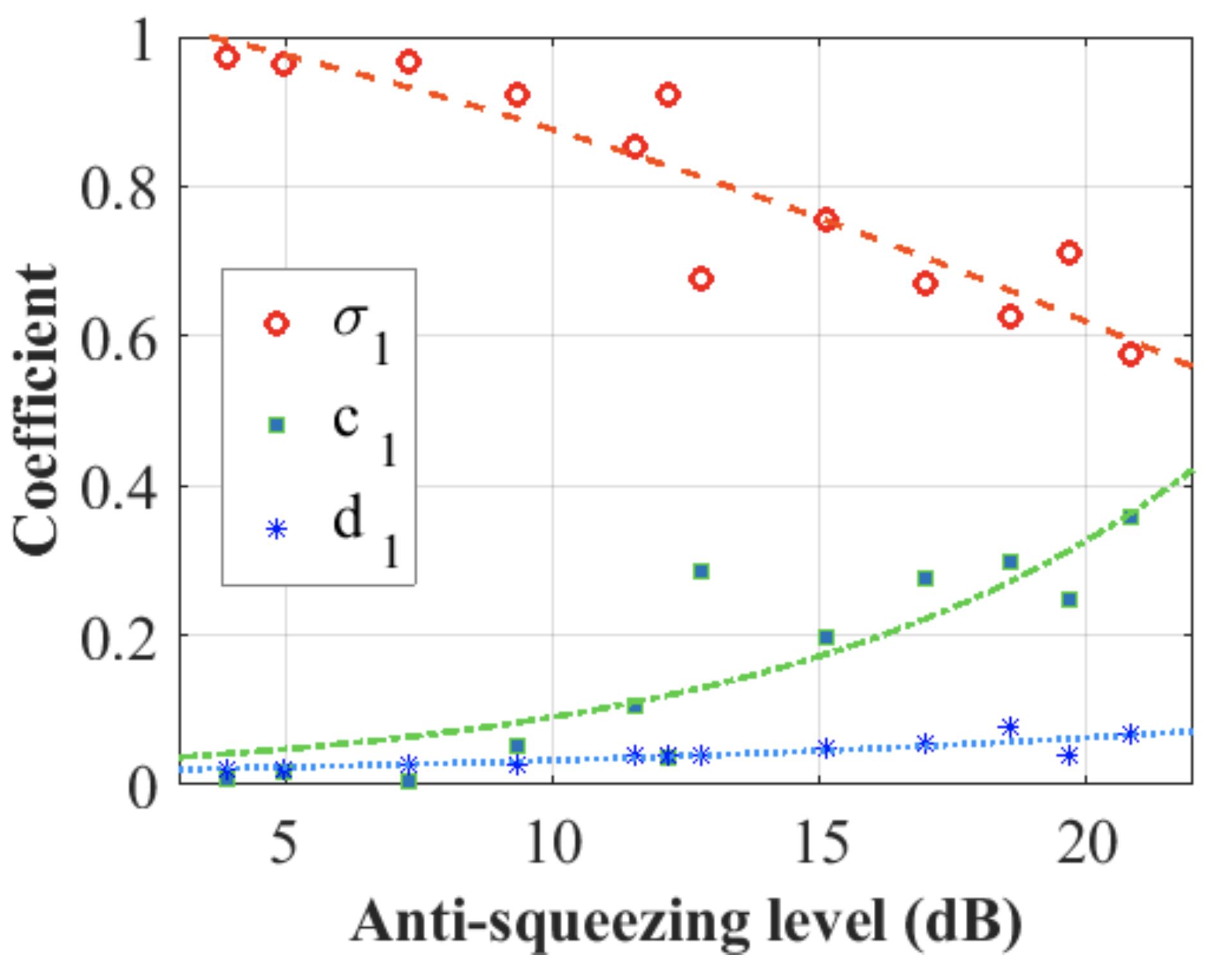}
\caption{With the help of machine learning, we can directly extract the degradation information from the obtained density matrix. Here, the three largest singular values correspond to the coefficients in ideal (pure) squeezed state, the squeezed thermal state, and thermal state, i.e., $\rho = \sigma_1\, \rho^{sq}+ c_1\, \rho^{sq}_{th} + d_1\, \rho_{th}$, respectively.}
\end{figure}

Furthermore, we can directly apply the singular value decomposition to the predicted density matrix and extract the dominant terms, i.e., $\rho = \sigma_1\, \rho^{sq}+ \sigma_{non}\, \rho^{non}$, 
where $\sigma_{non}\, \rho^{non} =  \sum_i c_i\, \rho^{sq}_{th, i} + \sum_i d_i\, \rho_{th, i}$ denotes the summation of all the contributed squeezed thermal states $\rho^{sq}_{th, i}$ and non-squeezed thermal states $\rho_{th, i}$, with the corresponding singular values $c_i$ and $d_i$, respectively.
For the four selected experimental data, we have $\sigma_1=0.9764$, $\sigma_{non}= 0.0236$ for the nearly idea squeezing (marker $A$); $\sigma_1=0.8568$, $\sigma_{non}= 0.1432$ and $\sigma_1=0.7109$, $\sigma_{non}= 0.289$ for the high squeezing level but with low (marker $B$) and high (marker $C$)  degradations, respectively. 
As for the highest squeezing level (markers $D$), we have $\sigma_1=0.5142$, $\sigma_{non}= 0.4858$.

 To precisely identify the pure squeezed and noisy parts, in Fig. 6, with the help of machine learning, we can directly extract the three largest singular values corresponding to the coefficients in ideal (pure) squeezed state, the squeezed thermal state, and thermal state, i.e.,$\rho = \sigma_1\, \rho^{sq}+ c_1\, \rho^{sq}_{th} + d_1\, \rho_{th}$, respectively.
Now, it clearly illustrates that there are two dominant terms in the degradation: from the contributions of  thermal and squeezed thermal states. As one expects, the thermal part, described by the coefficient $d_1$ in Blue-color, remains almost constant. It manifests the lossy effect due to the environment.
It is a common belief that  as long as the system is stable, the loss and phase noises can be measured by injecting classical laser light and be estimated.
On the contrary, the other lossy effect  from the squeezed thermal states, described by the coefficient $c_1$ in Green-color,  increases as the (anti-) squeezing increases. 
It is this unexpected squeezed thermal state causes the severe degradation at higher squeezing levels~\cite{sqthermal}, which demonstrates the advantages in applying machine learning to QST.

In conclusion, a neural network enhanced quantum state tomography is implemented experimentally for continuous variables. 
In particular, our well-trained machine fed with  squeezed vacuum  and squeezed thermal states not only completes the task of the reconstruction of  Wigner function in less than one second, but also keeps the high fidelity in the predict density matrix.
Compared to the over-estimation by MLE and under-estimation by empirically fitting at high squeezing levels, the purity of squeezed states at  squeezing level close to  $10$~dB is demonstrated experimentally, along with low and high anti-squeezing levels.
Such a fast, robust, and precise quantum state tomography enables us to extract the degradation information in squeezing  only with a single shot measurement.
Our experimental  implementations also act as the crucial diagnostic toolbox for the applications with squeezed states, including the advanced gravitational wave detectors, quantum metrology,  macroscopic quantum state generation, and quantum information process. 
In addition to the squeezed states illustrated here, similar concepts demonstrated in our well-trained machine can be readily applied to a specific family of continuous variables, such as non-Gaussian states.
Of course, different training (learning) processes should be applied in dealing with single-photon states, Cat states, and GKP states.
As we illustrated in this work, a supervised machine-learning, such as the CNN used here, provides a good starting point to implement QST with machine learning. In addition to CNN, with  a better kernel developed in machine-learnings, it is possible to use less training data with a variety of machine learning architectures. For example,  by applying reinforce learning~\cite{QML-3}, quantum machine learning is expected to provide an efficient and robust way to explore the quantum world.

\section*{Acknowledgement}
We are indebted to the helps from Prof. Akira Furusawa, Prof. Ping Koy Lam and Dr. Syed Assad. This work is partially supported by the Ministry of Science and Technology of Taiwan (No. 108-2923-M-007-001-MY3 and No. 109-2112-M-007-019-MY3), Office of Naval Research Global, and the collaborative research program of the Institute for Cosmic Ray Research (ICRR), the University of Tokyo.

\end{document}